\documentclass[prd,twocolumn,showpacs,preprintnumbers,amsmath,amssymb,superscriptaddress,floatfix,nofootinbib]{revtex4}

\usepackage{graphicx}
\usepackage{amsmath}
\usepackage{amsfonts}
\usepackage{amssymb}
\usepackage{color}
\usepackage{multirow}
\usepackage[colorlinks, citecolor=blue,anchorcolor=red,menucolor=red, linkcolor=red,filecolor=red,runcolor=red,urlcolor=blue,frenchlinks=red]{hyperref}

\begin{document}

\title{Canonical interpretation of the $X(4140)$ state within the $^3P_0$ model}

\author{Wei Hao}
\affiliation{School of Physics and Microelectronics, Zhengzhou University, Zhengzhou, Henan 450001, China}

\author{Guan-Ying Wang}
\affiliation{School of Physics and Microelectronics, Zhengzhou University, Zhengzhou, Henan 450001, China}

\author{En Wang} \email{wangen@zzu.edu.cn}
\affiliation{School of Physics and Microelectronics, Zhengzhou University, Zhengzhou, Henan 450001, China}

\author{Guan-Nan Li}
\affiliation{School of Physics and Microelectronics, Zhengzhou University, Zhengzhou, Henan 450001, China}

\author{De-Min Li}
\affiliation{School of Physics and Microelectronics, Zhengzhou University, Zhengzhou, Henan 450001, China}

\begin{abstract}
Recently, the LHCb Collaboration has confirmed the state $X(4140)$, with a mass $M=4146.5\pm 4.5^{+4.6}_{-2.8}$~MeV, and a much larger width $\Gamma=83\pm21^{+21}_{-14}$~MeV than the previous experimental measurements, which has confused the understanding of its nature. We will investigate the possibility of the $\chi_{c1}(3P)$ interpretation for the $X(4140)$, considering the mass spectra predicted in the quark model, and the strong decay properties within the $^3P_0$ model.  We also predict the strong decay properties of the charmonium states $\chi_{c0}(3P)$ and $\chi_{c2}(3P)$. Our results show that the $X(4140)$ state with the small width given in PDG can be explained as the charmonium state $\chi_{c1}(3P)$ in the $^3P_0$ model, and high precision measurement of the width of the $X(4140)$ is crucial to understand its nature.
\end{abstract}

\maketitle

\section{Introduction}
Since the $X(3872)$ was discovered in 2003 by the Belle Collaboration~\cite{Choi:2003ue}, a lot of unexpected states (charmonium-like states or XYZ states) have been reported experimentally~\cite{PDG2018}. Most of them have strange properties, and are difficult to be interpreted as the charmonium states, which makes them more like exotic states~\cite{Brambilla:2019esw,Chen:2016qju,Liu:2019zoy,Guo:2017jvc}.

In 2009, a new-threshold $X(4140)$ state was first reported in the $B^+ \to J/\psi \phi K^+$ process by the CDF Collaboration~\cite{Aaltonen:2009tz}, with a statistical significance of the signal $3.8\sigma$. This state was confirmed in the same process by the CMS~\cite{Chatrchyan:2013dma} and D0 Collaborations~\cite{Abazov:2013xda,Abazov:2015sxa}, and also in the reanalyzed the $B^\pm \to J/\psi \phi K^\pm$ process with a larger data sample by the CDF Collaboration\cite{Aaltonen:2011at}.
However, the Belle, LHCb, and Babar Collaborations have not found the signal of this state~\cite{Shen:2009vs,Aaij:2012pz,Lees:2014lra}. Since the $X(4140)$ is only seen in the $J/\psi \phi$ channel, which is OZI suppressed for the charmonium assignment, the hidden charm decay of this state disfavors the explanation of the charmonium $\chi_{cJ}(3P)$~\cite{Liu:2009iw}. There are a lot of theoretical interests about its properties, such as charmonium state, molecular state, tetraquark state, hybrid state, or a rescattering effect (more information can be found in the reviews~\cite{Chen:2016qju,Guo:2017jvc}).

In 2017, the LHCb Collaboration has also confirmed this state with high statistic data~\cite{Aaij:2016iza,Aaij:2016nsc}~\footnote{It should be stressed that LHCb have preformed twice analyses of the reaction $B^+\to J/\psi \phi K^+$ respectively in 2012 and 2017, and the first analysis of Ref.~\cite{Aaij:2012pz} has not found the evidence of $X(4140)$.}  with a mass $4146.5\pm 4.5^{+4.6}_{-2.8}$~MeV and a width $83\pm 21^{+21}_{-14}$~MeV, much larger than the previous experimental measurements (see the Table~\ref{tab:4140}), and the quantum numbers of this state were determined to be $J^{PC}=1^{++}$.
Thus, the $D^*_s\bar{D}^*_s$ molecular explanation, which prefers the quantum numbers $J^{PC}=0^{++}$ or $2^{++}$, is ruled out~\cite{Liu:2009ei,Branz:2009yt,Chen:2015fdn,Karliner:2016ith,Albuquerque:2009ak,Zhang:2009st,Wang:2009ue,Ding:2009vd}.

However, the $X(4140)$ is still the subject of much theoretical work, and there are many different suggestions about its structure~\cite{Chen:2016oma,Lu:2016cwr,Chen:2016iua,Wang:2018qpe,Wu:2016gas,Agaev:2017foq}. For instance, Ref.~\cite{Lu:2016cwr} regards the $X(4140)$ as the $cs\bar{c}\bar{s}$ tetraquark ground state. The $X(4140)$ state with the assignment of the $\chi_{c1}(3P)$ state is predicted to have a small width in Ref.~\cite{Chen:2016iua}. In Ref.~\cite{Wang:2018qpe}, the partial width of the decay mode $X(4140)\to J/\psi\phi$ is predicted to be $86.9\pm22.6$~MeV, with the axial-vector tetraquark picture for the $X(4140)$. In addition, the width of the $X(4140)$ is predicted to be $80\pm29$~MeV with in the interpretation of the color triplet diquark-antidiquark state~\cite{Agaev:2017foq}, and Refs.~\cite{Liu:2016onn,Ortega:2016hde} have claimed that the structure of the $X(4140)$ may be the cusp due to the presence of the $D_s^{*+}D_s^-$ (or $D_s^{*-}D_s^+$) threshold.
Recently, Ref.~\cite{Turkan:2017pil} points out that it is not possible to claim the molecular or diquark-antidiquark content of the $X(4140)$ within the QCD sum rules.

Indeed, it is natural and necessary to exhaust the possible $q\bar{q}$ description of the observed states before restoring to the more exotic assignments. While the ground states of the $P$-wave charmonium states, $\chi_{cJ}(1P)$, have been well established, and the first radial excitations, $\chi_{cJ}(2P)$, are predicted to have the mass around 3900~MeV~\cite{PDG2018,Guo:2012tv,Liu:2012ze,Li:2009zu,Olsen:2014maa,Wang:2014voa},  the $X(4140)$, with the quantum numbers of $J^{PC}=1^{++}$, could be the second radial excitation $\chi_{c1}(3P)$, with the predicted mass of $4100\sim 4200$~MeV in the quark model~\cite{Li:2009zu,Wang:2019mhs}. It should be
noted that the mass information alone is insufficient to
classify the $X(4140)$, so its decay behaviors also
need to be compared with model expectations.

 In this work, taking the meson wave functions obtained from the relativistic/non-relativistic quark models, we will investigate the decay properties of the $X(4140)$ as the assignment of charmonium state in the $^3P_0$ model, and provide more information about the decay modes, since the observation of the $X(4140)$ in other channels could be useful to extract its width with more precisely.

\begin{table}[htpb]
\begin{center}
\caption{ \label{tab:4140} The experimental measurements of the $X(4140)$ (in MeV).}
\footnotesize
\setlength{\tabcolsep}{1mm}{
\begin{tabular}{ccccc}
\hline\hline
  Exp.                 &Mass    &Width      & Sig.         &Year    \\\hline
  CDF~\cite{Aaltonen:2009tz}      &$4143.0\pm 2.9\pm 1.2$               &$11.7^{+8.3}_{-5.0}\pm3.7$ & 3.8$\sigma$       &2009                 \\
  CMS~\cite{Chatrchyan:2013dma}      &$4148.0\pm 2.4\pm6.3 $               &$28^{+15}_{-11}\pm19$           &  5.0$\sigma$          &2014                     \\

  D0~\cite{Abazov:2013xda}      &$4159.0\pm4.3\pm6.6$                 &$20\pm13^{+3}_{-8} $  & 3.0$\sigma$                     &2014                       \\
  D0~\cite{Abazov:2015sxa}      &$4152.5\pm1.7^{+6.2}_{-5.4}$         &$16.3\pm5.6\pm11.4$   & 4.7$\sigma$                         &2015                      \\
  CDF~\cite{Aaltonen:2011at}      &$4143.4^{+2.9}_{-3.0}\pm0.6$         &$15.3^{+10.4}_{-6.1}\pm2.5$ & 5.0$\sigma$         &2011                 \\
  LHCb~\cite{Aaij:2016nsc}    &$4146.5\pm4.5^{+4.6}_{-2.8}$         &$83\pm21^{+21}_{-14}$          & 8.4$\sigma$ &2017                       \\
  PDG~\cite{PDG2018} &  $4146.8\pm 2.4 $ & $22^{+8}_{-7}$ & & 2019 \\
  \hline\hline
\end{tabular}}
\end{center}
\end{table}


This paper is organized as follows. In Sec.~\ref{sec:model}, we will present a brief review of the $^3P_0$ decay model, and in Sec.~\ref{sec:wavefunctions}, we will introduce two kinds of wave functions for the mesons. The results and  discussions  are shown in Sec.~\ref{sec:results}. Finally, the summary is given in Sec.~\ref{sec:summary}.

\section{The $^3P_0$ decay model}
\label{sec:model}

\begin{figure}[htpb]
  \centering
  \includegraphics[scale=0.4]{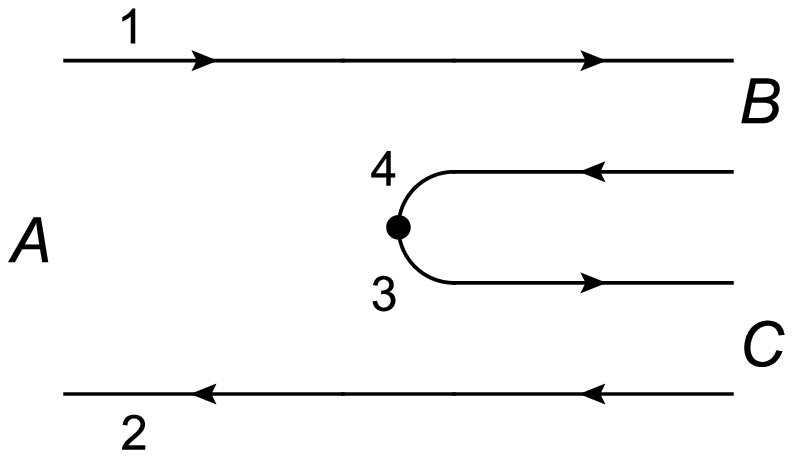}
    \includegraphics[scale=0.4]{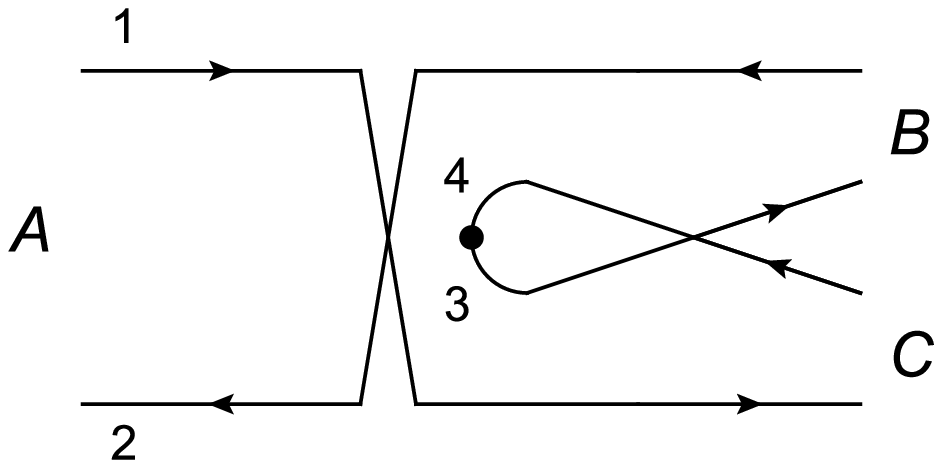}
  \caption{The two possible diagrams contributing to $A \to BC$ in
the $^3P_0$ model: (left) the quark within the meson $A$ combines with
the created antiquark to form the meson $B$, the antiquark within
the meson $A$ combines with the created quark to form the meson
$C$; (right) the quark within the meson $A$ combines with the created
antiquark to form the meson $C$, the antiquark within the meson $A$
combines with the created quark to form the meson $B$.}
  \label{sec:3P0}
\end{figure}

In this section, we will present the $^3P_0$ model, which is used to evaluate the Okubo-Zweig-Iizuka (OZI) allowed open charm decays of the $\chi_{cJ}(3P)$.
The $^3P_0$ model, also known as the quark-pair creation model, was originally introduced by Micu~\cite{Micu:1968mk} and further developed by Le Yaouanc $et$ $al.$~\cite{LeYaouanc:1972vsx,LeYaouanc:1973ldf,LeYaouanc:alo}. The $^3P_0$ model has been widely applied to study strong decays of hadrons with considerable success~\cite{Roberts:1992js,Blundell:1996as,Barnes:1996ff,Barnes:2002mu,
Close:2005se,Barnes:2005pb,Zhang:2006yj,Ding:2007pc,Li:2008mza,Li:2008we,Li:2008et,Xue:2018jvi,
Li:2009rka,Li:2009qu,Li:2010vx,Wang:2017pxm,Pan:2016bac,Lu:2016bbk}.
In this model, the strong decay of hadron occurs through a quark-antiquark pair created from the vacuum with the vacuum quantum number $J^{PC}=0^{++}$, then the new quark-antiquark pair, together with the $q\bar{q}$ within the initial meson, regroups into two outgoing mesons in all possible quark rearrangement ways, as shown in Fig.~\ref{sec:3P0}~\footnote{It should be pointed out that these two diagrams of Fig.~\ref{sec:3P0} are different, and they will give flavor weight factors for a specified flavor channel~\cite{Barnes:1996ff}. For instance, the process of $\rho^+$ (A) decay to $\pi^+$ (B) and $\pi^0$(c) can perform as left diagram by creating $\bar{d}d$ quark pair, and also can perform as right diagram by creating $\bar{u}u$ pair. }.

The transition operator $T$ of the decay $A\rightarrow BC$ in the $^3P_0$ model can be written,
\begin{eqnarray}
  T &=&-3\gamma\sum_m\langle 1m1-m|00\rangle\int
    d^3\boldsymbol{p}_3d^3\boldsymbol{p}_4\delta^3(\boldsymbol{p}_3+\boldsymbol{p}_4)\nonumber\\
&& \times   {\cal{Y}}^m_1\left(\frac{\boldsymbol{p}_3-\boldsymbol{p}_4}{2}\right
    )\chi^{34}_{1,-m}\phi^{34}_0\omega^{34}_0b^\dagger_3(\boldsymbol{p}_3)d^\dagger_4(\boldsymbol{p}_4),
\end{eqnarray}
where $\gamma$ is a dimensionless parameter corresponding to the strength of quark-antiquark $q_3\bar{q}_4$ pair produced from the vacuum, and $\boldsymbol{p}_3$ and $\boldsymbol{p}_4$ are the momenta of the created quark $q_3$ and antiquark $\bar{q}_4$, respectively. $\chi^{34}_{1,-m}$, $\phi^{34}_0$, and $\omega^{34}_0$ are the spin, flavor, and color wave functions of the $q_3\bar{q}_4$, respectively. The solid harmonic polynomial  ${\cal{Y}}^m_1(\boldsymbol{p})\equiv|p|^1Y^m_1(\theta_p, \phi_p)$ reflects the momentum-space distribution of the $q_3\bar{q}_4$.

The partial wave amplitude ${\cal{M}}^{LS}(\boldsymbol{P})$ of the decay  $A\rightarrow BC$ can be given by~\cite{Jacob:1959at},
\begin{eqnarray}
{\cal{M}}^{LS}(\boldsymbol{P})&=&
\sum_{\renewcommand{\arraystretch}{.5}\begin{array}[t]{l}
\scriptstyle M_{J_B},M_{J_C},\\\scriptstyle M_S,M_L
\end{array}}\renewcommand{\arraystretch}{1}\!\!
\langle LM_LSM_S|J_AM_{J_A}\rangle \nonumber\\
&&\langle
J_BM_{J_B}J_CM_{J_C}|SM_S\rangle\nonumber\\
&&\times\int
d\Omega\,Y^\ast_{LM_L}{\cal{M}}^{M_{J_A}M_{J_B}M_{J_C}}
(\boldsymbol{P}), \label{pwave}
\end{eqnarray}
where ${\cal{M}}^{M_{J_A}M_{J_B}M_{J_C}}
(\boldsymbol{P})$ is the helicity amplitude and defined as,
\begin{eqnarray}
\langle
BC|T|A\rangle=\delta^3(\boldsymbol{P}_A-\boldsymbol{P}_B-\boldsymbol{P}_C){\cal{M}}^{M_{J_A}M_{J_B}M_{J_C}}(\boldsymbol{P}).
\end{eqnarray}
The $|A\rangle$, $|B\rangle$, and $|C\rangle$ denote the mock meson states defined by Ref.~\cite{Hayne:1981zy}.

Due to different choices of the pair-production vertex, phase space convention, employed meson space wave function, various $^3P_0$ models exist in literature. In this work, we employ the simplest vertex as introduced originally by Micu which assumes a spatially  constant pair-production strength $\gamma$~\cite{Micu:1968mk}, and adopt the relativistic phase space. We will take into account the two choices of the wave functions for mesons, which will be presented in next section.
Finally, the decay width
$\Gamma(A\rightarrow BC)$ can be expressed in terms of the partial wave amplitude,
\begin{eqnarray}
\Gamma(A\rightarrow BC)= \frac{\pi
|\boldsymbol{P}|}{4M^2_A}\sum_{LS}|{\cal{M}}^{LS}(\boldsymbol{P})|^2, \label{width1}
\end{eqnarray}
where $|\boldsymbol{P}|=\frac{\sqrt{[M^2_A-(M_B+M_C)^2][M^2_A-(M_B-M_C)^2]}}{2M_A}$,
and $M_A$, $M_B$, and $M_C$ are the masses of the meson $A$, $B$, and $C$, respectively. The explicit expressions for  ${\cal{M}}^{LS}(\boldsymbol{P})$ can be found in Refs.~\cite{Li:2008mza,Li:2008we,Li:2008et}.

\section{Wave functions}
\label{sec:wavefunctions}
In this section, we will present two choices of wave functions for the charmonium states, charm and charmed-strange mesons, which will be used to calculate the $\chi_{cJ}(3P)$ strong decay widths.

As discussed in Ref.~\cite{Li:2009zu}, the quenched quark models, which incorporates a coulomb term at short distances and the linear confining interaction at large distances, will not be reliable in the domain beyond the open-charm threshold. This is because the linear potential, which is expected to be dominant in this mass region, will be screened or softened by the vacuum polarization effects of dynamical fermions. We will adopt the  Godfrey-Isgur model and Non-relativistic quark model, modified to incorporate the screening potential to account for the screening effects. In the following, we will see that both the modified Godfrey-Isgur model and modified Non-relativistic quark model could provide a nice description for the high excited charmonium states.

\begin{table*}[htpb]
\begin{center}
\caption{ \label{tab:d} The mass spectra (in MeV) of charm mesons obtained within the non-relativistic quark model (NRQM) and the modified Godfrey-Isgur Model (MGI).}
\footnotesize
\begin{tabular}{ccccc}
\hline\hline
   $n^{2S+1}L_J$  & states                    &PDG\cite{PDG2018}        &NRQM~\cite{Li:2010vx}    &MGI\cite{Song:2015fha}               \\\hline
  $(1^1S_0)$    & $D$           &$1864.83\pm0.05/1869.65\pm0.05$     &1867   & 1861          \\
  $(1^3S_1)$    & $D^*$         &$2006.85\pm0.05/2010.26\pm0.05$     &2010   & 2020         \\
  $(2^1S_0)$   & $D(2550)$      &$2564\pm20$                         &2555   & 2534        \\
  $(2^3S_1)$ &               &                                    &2636   & 2593         \\
  $(3^1S_0)$  &              &                                    &3047   & 2976       \\
  $(3^3S_1)$  &              &                                    &3109   & 3015         \\
  $(4^1S_0)$  &              &                                    &3464   & 3326         \\
  $(4^3S_1)$  &              &                                    &3516   & 3353          \\
  $(1P)$  & $D_1(2420)$     &$2420.8\pm0.5$                      &2402   & 2426        \\
  $(1^3P_0)$ & $D_{0}^{*}(2400)$ &$2300\pm19$                         &2252   & 2365       \\
  $(1P^\prime)$ & $D_1(2430)$      &$2427\pm40$                         &2417   & 2431      \\
  $(1^3P_2)$ & $D_{2}^*(2460)$  &$2460.7\pm0.4/2465.4\pm1.3$         &2466   & 2468     \\
  $(2P)$ &               &                                    &2886   & 2861        \\
  $(2^3P_0)$ &               &                                    &2752   & 2856      \\
  $(2P^\prime)$ &               &                                    &2929   & 2877           \\
  $(2^3P_2)$ &               &                                    &2971   & 2884           \\
  $(1D)$ &              &                                     &2693  & 2773           \\
  $(1^3D_1)$ &               &                                    &2740   & 2762         \\
  $(1D^\prime)$ &               &                                    &2789   & 2779         \\
  $(1^3D_3)$ & $D_{3}^*(2750)$  &$2763.5\pm3.4$                  &2719   & 2779                      \\
  $(2D)$ &               &                                    &3145   & 3128         \\
  $(2^3D_1)$ &               &                                    &3168  & 3131     \\
  $(2D^\prime)$ &               &                                    &3215   & 3136         \\
  $(2^3D_3)$ &               &                                    &3170   & 3129          \\
  \hline\hline

\end{tabular}
\end{center}
\end{table*}

\begin{table}[htpb]
\begin{center}
\caption{ \label{tab:ds}The mass spectra (in MeV) of charmed-strange mesons obtained within the non-relativistic quark model (NRQM) and the modified Godfrey-Isgur Model (MGI). }
\footnotesize
\begin{tabular}{ccccc}
\hline\hline
  $n^{2S+1}L_J$  & states                             & PDG~\cite{PDG2018}  &NRQM~\cite{Li:2010vx}     &MGI~\cite{Song:2015nia}               \\\hline
  $(1^1S_0)$ & $D_{s}$              &$1968.34\pm0.07$                    &1969   & 1967           \\
  $(1^3S_1)$ & $D_{s}^{*}$           &$2112.2\pm0.4$                      &2107   & 2115                \\
  $(2^1S_0)$ &                       &                                    &2640   & 2646               \\
  $(2^3S_1)$ & $D_{s1}^*(2700)$     &$2708.3_{-3.4}^{+4.0}$              &2714   & 2704                \\
  $(3^1S_0)$ &                       &                                    &3112   & 3097                \\
  $(3^3S_1)$ &                       &                                    &3168   & 3136             \\
  $(4^1S_0)$ &                       &                                    &3511   & 3462                         \\
  $(4^3S_1)$ &                       &                                    &3558   & 3490                          \\
  $(1P)$ & $D_{s1}(2536)$       &$2535.11\pm0.06$                    &2488   & 2531          \\
  $(1^3P_0)$ & $D_{s0}^*(2317)$     &$2317.8\pm0.5$                      &2344   & 2463           \\
  $(1P^\prime)$ & $D_{s1}(2460)$       &$2459.5\pm0.6$                      &2510   & 2532            \\
  $(1^3P_2)$ & $D_{s2}^*(2573)$      &$2569.1\pm0.8$                      &2559   & 2571          \\
  $(2P)$ &                       &                                    &2958   & 2979            \\
  $(2^3P_0)$ &                       &                                    &2830   & 2960            \\
  $(2P^\prime)$ &                       &                                    &2995   & 2988                 \\
  $(2^3P_2)$ &                       &                                    &3040   & 3004                 \\
  $(1D)$ &                       &                                    &2788  & 2877                \\
  $(1^3D_1)$ & $D_{s1}^*(2860)$     &$2859\pm27$                         &2804   &2865               \\
  $(1D^\prime)$ &                       &                                    &2849   & 2882               \\
  $(1^3D_3)$ & $D_{s3}^*(2860)$     &$2860\pm7$                          &2811   & 2883                        \\
  $(2D)$ &                       &                                    &3217   & 3247               \\
  $(2^3D_1)$ &                       &                                    &3217  & 3244               \\
  $(2D^\prime)$ &                       &                                    &3260   &3252                \\
  $(2^3D_3)$ &                       &                                    &3240   & 3251                 \\
  \hline\hline

\end{tabular}
\end{center}
\end{table}

\subsection{Non-relativistic quark model}
\label{sec:NRQM}
For the wave functions of the open charm mesons in the final states, we use the non-relativistic quark model (NRQM), proposed by Lakhina and Swanson~\cite{Lakhina}. This non-relativistic quark model has been successfully used to describe the mass spectrum of charm and charmed-strange mesons~\cite{Lakhina,Li:2010vx}, bottom mesons~\cite{Lu:2016bbk}.

For the open charm mesons, the total Hamiltonian can be written as~\cite{Li:2009zu}
\begin{equation}
    H = H_0+H_{sd}+C_{q\bar{q}},\label{eq:hamiltonA}
    \end{equation}
where $H_0$ is the zeroth-order Hamiltonian, $H_{sd}$ is the spin-dependent Hamiltonian, and $C_{q\bar{q}}$ is a constant.
The $H_{0}$ can be compressed as
    \begin{equation}
      H_0 = \frac{\boldsymbol{p}^2}{M_r}-\frac{4}{3}\frac{{\alpha}_s}{r}+br,
    \end{equation}
where $\boldsymbol{p}$ is the center-of-mass momentum, $r$ is the $q\bar{q}$ separation, $M_r=2m_qm_{\bar{q}}/(m_q+m_{\bar{q}})$,  $m_q$ and $m_{\bar{q}}$ are the masses of quark $q$ and anti-quark $\bar{q}$, respectively, $b=0.14$~GeV$^2$ is the linear potential slope and $\alpha_s=0.5$ is the coefficient of Coulomb potential~\cite{Lakhina,Li:2010vx}.
The explicit expression of the $H_{sd}$ and the corresponding parameters are given in Refs.~\cite{Lakhina,Li:2010vx}. We have tabulated the spectra of charm and charmed-strange mesons in Table~\ref{tab:d} and Table~\ref{tab:ds}, respectively, which are same as those of Ref.~\cite{Li:2010vx}.

For the wave functions of the charmonium states, we will use the modified non-relativistic quark model (MNRQM) by taking into account the screening effect, as discussed in Ref.~\cite{Li:2009zu}.
When screening effect is considered, the modification can be accomplished by the  transformation
\begin{equation}
br\to V^{scr}(r)= \frac{b(1-e^{\mu r})}{\mu},
\end{equation}
where $\mu=0.0979$~GeV is the characteristic scale for color screening, and $b=0.21$~GeV$^2$~\cite{Li:2009zu}. The mass spectra of the charmonium states are shown in Table~\ref{tab:ccbar}, which are same as those of Ref.~\cite{Li:2009zu}.

\subsection{Modified Godfrey-Isgur model}
\label{sec:MGI}
In addition to the non-relativistic quark model, the Godfrey-Isgur (GI) relativistic quark model~\cite{Godfrey:1985xj} is one of the most successful models describing mass spectrum of mesons. Because the coupled-channel effect becomes more important for higher radial and orbital excitations, the modified relativistic quark model was proposed \cite{Song:2015fha,Song:2015nia}
and widely used to calculate mass spectrum of charm meson~\cite{Song:2015fha}, charmed-strange meson~\cite{Song:2015nia}, charmonium~\cite{Wang:2019mhs} and  bottomonium~\cite{Wang:2018rjg}.
In the relativistic quark model, the Hamiltonian of a meson system is \cite{Godfrey:1985xj}
\begin{eqnarray}
\tilde{H}&=&\left(p^2+m_q^2\right)^{1/2}+\left(p^2+m_{\bar{q}}^2\right)^{1/2}\\
&&+\tilde{H}_{q\bar{q}}^{\text{conf}}+\tilde{H}_{q\bar{q}}^{\text{so}}+\tilde{H}_{q\bar{q}}^{\text{hyp}}. \label{eq:hamiltonB}
\end{eqnarray}
where $\tilde{H}^{conf}_{q\bar{q}}$ is spin-independent potential, $\tilde{H}^{hyp}_{q\bar{q}}$ is color-hyperfine interaction, $\tilde{H}^{so}_{q\bar{q}}$ is spin-orbit interaction. The explicit expression of $\tilde{H}^{conf}_{q\bar{q}}$, $\tilde{H}^{hyp}_{q\bar{q}}$, and $\tilde{H}^{so}_{q\bar{q}}$ are given in Ref.~\cite{Song:2015nia}. The spin-independent potential contains a constant term, a linear confining potential, and a one-gluon exchange potential,
\begin{equation}
\tilde{H}^{conf}_{q\bar{q}}=c + b r +\frac{\alpha_s(r)}{r} F_1 \cdot F_2. \label{eq:hconf}
\end{equation}

Although the GI model has achieved great successes in describing the meson spectrum, there still exists a discrepancy between the predictions and the recent experimental observation, as discussed in Refs.~\cite{Song:2015nia,Wang:2019mhs}.
When screening effect is considered, the modification can be accomplished by the  transformation~\cite{Wang:2019mhs}
\begin{equation}
br\to V^{scr}(r)= \frac{b(1-e^{\mu r})}{\mu},
\end{equation}
where the $b=0.2687$~GeV$^2$ and $\mu=0.15$~GeV\cite{Wang:2019mhs}.

With the modified Godfrey-Isgur (MGI) model, we calculated the mass spectra of charm mesons, charmed-strange mesons, and charmonium states, as shown  in Table~\ref{tab:d}, \ref{tab:ds}, and \ref{tab:ccbar}, respectively, which are same as those of Refs.~\cite{Song:2015fha,Song:2015nia,Wang:2019mhs}.

\begin{table*}[htpb]
\begin{center}
\caption{ \label{tab:ccbar}The mass spectra (in MeV) of charmonium states obtained within the modified non-relativistic quark model (MNRQM) and the modified Godfrey-Isgur Model (MGI), and the other predictions are also listed in this table. }
\footnotesize
\begin{tabular}{lccccccccc}
\hline\hline
  states                & PDG~\cite{PDG2018}  & MNRQM~\cite{Li:2009zu} & MGI~\cite{Wang:2019mhs}&\cite{Barnes:2005pb} &\cite{Barnes:2005pb} &\cite{Radford:2007vd}   &\cite{Godfrey:1985xj} &\cite{Cao:2012du} &\cite{Segovia:2013wma}            \\\hline
  $\eta_c(1^1S_0)$      &$2983.9\pm0.5$           & 2979     & 2981    &2982   &2975   &2980.3      &2970    &2978.4     &2990      \\
  $J/\psi(1^3S_1)$      &$3096.9\pm0.006$         & 3097     & 3096    &3090   &3098   &3097.36     &3100    &3087.7     &3096          \\
  $\eta_c(2^1S_0)$      &$3637.5\pm1.1$            & 3623     & 3642    &3630   &3623   &3597.1      &3620    &3646.9     &3643        \\
  $\psi(2^3S_1)$        &$3686.097\pm0.025$       & 3673     & 3683    &3672   &3676   &3685.5      &3680    &3684.7     &3703                                \\
  $\eta_c(3^1S_0)$      &                          & 3991     & 4013    &4043   &4064   &4014.0      &4060    &4058.0     &4054                             \\
  $\psi(3^3S_1)$        &$4039\pm1$               & 4022     & 4035    &4072   &4100   &4094.9      &4100    &4087.0     &4097                                \\
  $\eta_c(4^1S_0)$      &                          & 4250     & 4260    &4384   &4425   &            &        &4391.4     &                              \\
  $\psi(4^3S_1)$        &                       & 4273     & 4274    &4406   &4450   &4433.3      &        &4411.4     &                             \\
  $h_c(1^1P_1)$         &$3525.38\pm0.11$          & 3519     & 3538    &3516   &3517   &3526.9      &3520    &3526.9     &3515                           \\
  $\chi_{c0}(1^3P_0)$   &$3414.71\pm0.30$         & 3433     & 3464    &3424   &3445   &3415.7      &3440    &3366.3     &3452                          \\
  $\chi_{c1}(1^3P_1)$   &$3510.67\pm0.05$         & 3510     & 3530    &3505   &3510   &3508.2      &3510    &3517.7     &3504                            \\
  $\chi_{c2}(1^3P_2)$   &$3556.17\pm0.07$       & 3554     & 3571    &3556   &3550   & 3557.7     &3550    &3559.3     &3532                               \\
  $h_c(2^1P_1)$         &                         & 3908     & 3933    &3934   &3956   & 3960.5     &3960    &3941.9     &3956                            \\
  $\chi_{c0}(2^3P_0)$   &                       & 3842     & 3896    &3852   &3916   &3843.7      &3920    &3842.7     &3909                          \\
  $\chi_{c1}(2^3P_1)$   &                        & 3901     & 3929    &3925   &3953   &3939.7      &3950    &3935.0     &3947                               \\
  $\chi_{c2}(2^3P_2)$   &$3927.2\pm2.6$          & 3937     & 3952    &3972   &3979   &3993.7      &3980    &3973.1     &3969                          \\
  $h_c(3^1P_1)$         &                         & 4184     & 4200    &4279   &4318   &            &        &4309.7     &4278                           \\
  $\chi_{c0}(3^3P_0)$   &                        & 4131     & 4177    &4202   &4292   &            &        &4207.6     &4242                             \\
  $\chi_{c1}(3^3P_1)$   &                        & 4178     & 4197    &4271   &4317   &            &        &4298.7     &4272                                 \\
  $\chi_{c2}(3^3P_2)$   &                        &  4208    & 4213    &4317   &4337   &            &        &4352.4     &                             \\
   $\psi(1^1D_2)$        &                        &3796      & 3848    &3799   &3837   &3823.6      &3840    &3815.1     & 3812                                 \\
  $\psi(1^3D_1)$        &$3773.13\pm0.35$         & 3787     & 3830    &3785   &3819   &3803.8      &3820    &3808.8     & 3796                          \\
  $\psi_2(1^3D_2)$      &                         & 3798     &3848     &3800   &3838   &3823.8      &3840    &3820.1     & 3810                          \\
  $\psi_3(1^3D_3)$      &                        &  3799    & 3859    &3806   &3849   &3831.1      &        &3812.6     &                            \\
  $\psi(2^1D_2)$        &                         & 4099     &4137     &4158   &4208   &4190.7      &4210    &4164.9     &4166                         \\
  $\psi(2^3D_1)$        &                          & 4089     &4125     &4142   &4194   &4164.2      &4190    &4154.4     &4153                          \\
  $\psi_2(2^3D_2)$      &                          & 4100     & 4137    &4158   &4208   &4189.1      &4210     &4168.7    &4160                        \\
  $\psi_3(2^3D_3)$      &                        & 4103     & 4144    &4167   &4217   &4202.3      &          &4166.1    &                       \\
  \hline\hline

\end{tabular}
\end{center}
\end{table*}

\section{Results and discussions}
\label{sec:results}

The mass spectra of the charmonium states predicted by the MNRQM and MGI models are shown in Table~\ref{tab:ccbar}. Taking into account the averaged mass of $X(4140)$,  $4146.8\pm 2.4$~MeV, and the quantum numbers of $I^G(J^{PC})=0^+(1^{++})$, we can tentatively assign the resonance $X(4140)$ as the candidate of the $\chi_{c1}(3P)$.
 The discrepancy between the averaged mass of $X(4140)$ and the predicted masses of $\chi_{c1}(3P)$ in both models maybe result from that the hadron loop effects (such as the $D\bar{D}$ loop), which were neglected in these two models. The hadron loop effects can give rise to mass shifts to the bare hadron states. The mass shifts induced by the hadron loop effects can present a better description of the $D$, $D_s$, charmonium states, and bottomonium states~\cite{Barnes:2007xu,Zhou:2011sp}.

Next, we will calculate the strong decay widths of the $X(4140)$ state as the $\chi_{c1}(3P)$  assignment.
In our calculations, we take two kinds of the wave functions, by solving the Schr\"{o}dinger equation in the (modified) NRQM as discussed in Subsec.~\ref{sec:NRQM} ({\bf Case A}), and in the MGI model as discussed in Subsec.~\ref{sec:MGI} ({\bf Case B}) for the charm mesons, charmed-strange mesons, and the charmonium states.  In the $^3P_0$ model, we take the same constituent quark masses as those in Eq.~(\ref{eq:hamiltonA}) for Case A ($m_{u/d}=450$~MeV and $m_s=550$~MeV), and as those in Eq.~(\ref{eq:hamiltonB}) for Case B ($m_{u/d}=220$~MeV and $m_s=419$~MeV).  Another free parameter $\gamma$,  the strength of quark-antiquark pair created from the vacuum, is taken to be $\gamma=4.52\pm 0.08$ in Case A,  and $\gamma=5.90\pm 0.10$ for Case B, by fitting to the total widths of the well established charmonium states,  $\psi(3770)$  ($1^3D_1$), $\psi(4040)$ ($3^3S_1$), $\psi(4160)$ ($2^3D_1$), and $\chi_{c2}(2P)$.

\begin{figure}[htpb]
  \centering
  \includegraphics[scale=0.6]{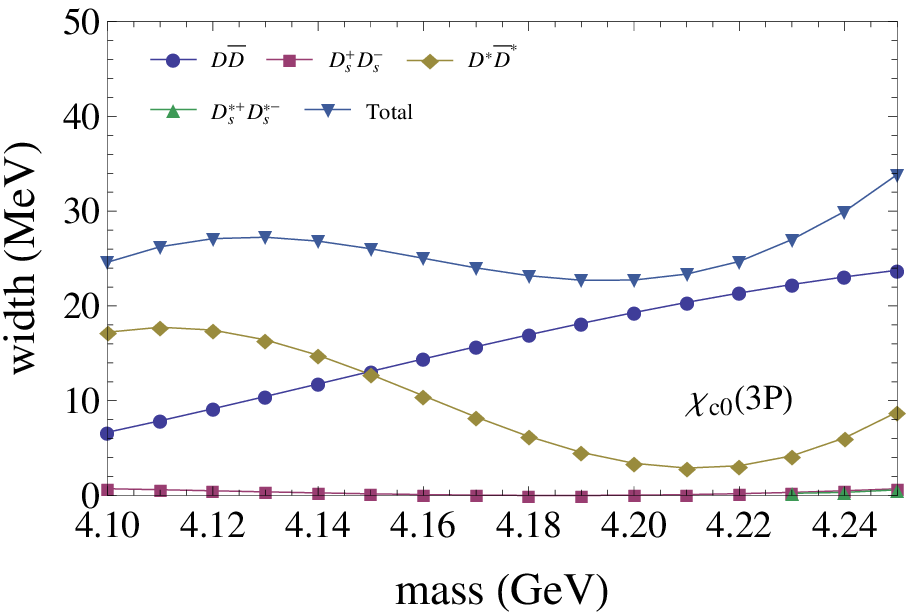}
  \includegraphics[scale=0.6]{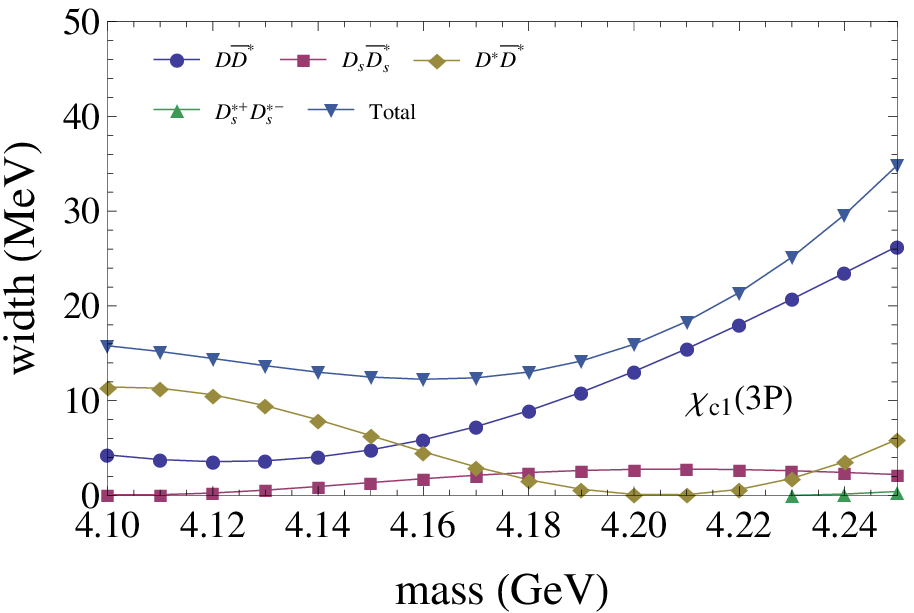}
  \includegraphics[scale=0.6]{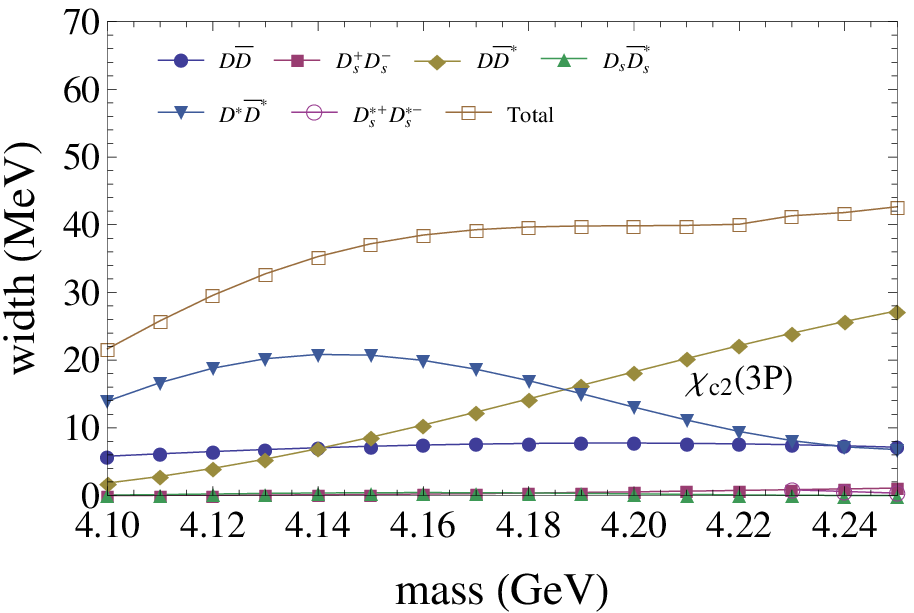}
  \caption{The dependences of the widths of ${\chi}_{c0}(3P)$, ${\chi}_{c1}(3P)$ and ${\chi}_{c2}(3P)$ on the initial state mass with the wave functions of Case A.}
  \label{fig:widthNRQM}
\end{figure}

\begin{figure}[htpb]
  \centering
\includegraphics[scale=0.6]{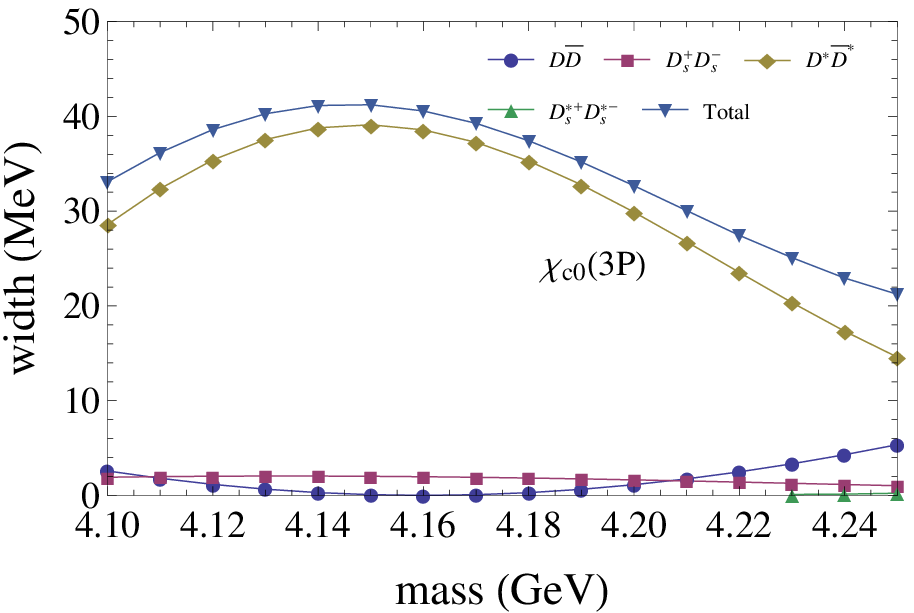}
\includegraphics[scale=0.6]{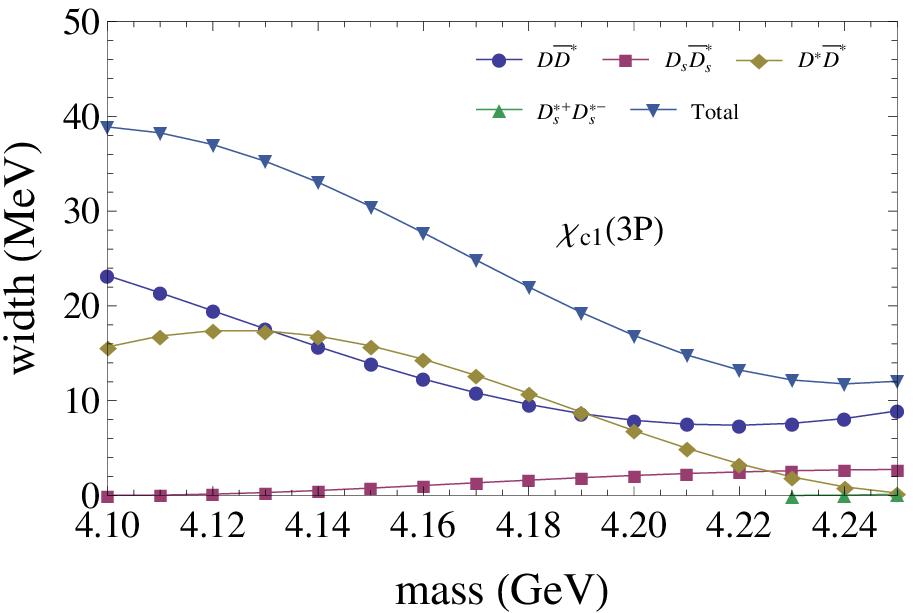}
\includegraphics[scale=0.6]{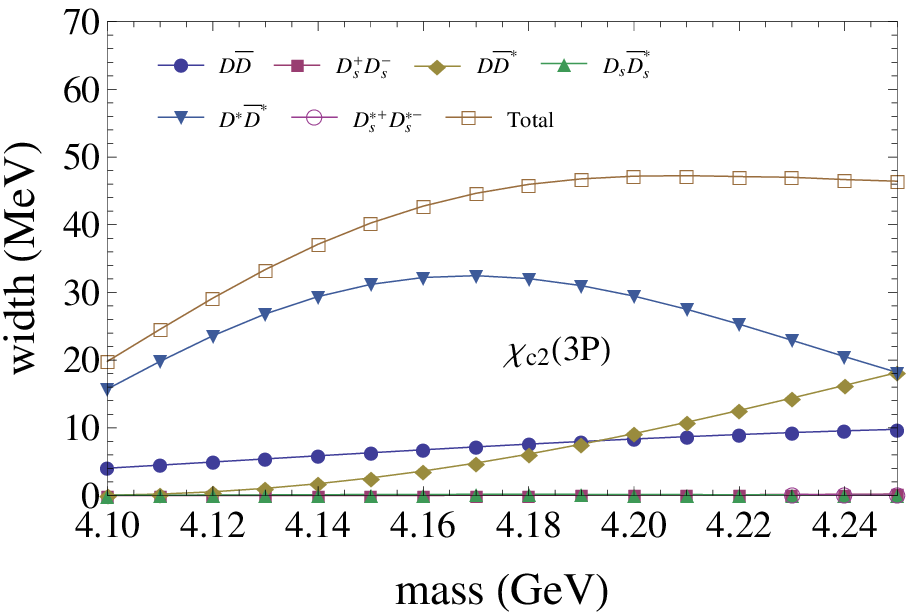}
  \caption{The dependences of the widths of ${\chi}_{c0}(3P)$, ${\chi}_{c1}(3P)$ and ${\chi}_{c2}(3P)$ on the initial state mass with the wave function of Case B.}
  \label{fig:widthMGI}
\end{figure}

With the above parameters, we have calculated the partial decay widths and total decay width, as shown in Table~\ref{tab:width4140} for both Case A and Case B. The total widths of $\chi_{c1}(3P)$ are 12.63~MeV for Case A, and $31.34$~MeV for Case B, both of which are consistent with the average value of $\Gamma=22^{+8}_{-7}$~MeV within errors~\cite{PDG2018}.  It should be pointed out that the decay modes $D\bar{D}^*$ and $D^*\bar{D}^*$ have large decay widths, which are also consistent with the conclusions of Refs.~\cite{Chen:2016iua,Wang:2014voa}. We suggest to search for this state in those two channels, and to measure the width precisely, which can be shed light on its nature.
We also show the dependence of the $\chi_{c1}(3P)$ decay width on the initial mass with the wave functions of Case A and Case B, respectively in Fig.~\ref{fig:widthNRQM} and Fig.~\ref{fig:widthMGI}. The decay width of the $\chi_{c1}(3P)$ state is $12.63\pm 0.45$~MeV for Case A, and $31.3\pm1.2$~MeV for Case B, by taking into account the uncertainties of the $X(4140)$ mass and the strength $\gamma$.

Since the error of the LHCb measurement on the width of $X(4140)$ is quite large, we will perform a simple $\chi^2$ study. For the results of the Case A, we have,\footnote{For the LHCb measurement, we use $83\pm30$~MeV by square summing the errors as $\sqrt{(21^2+21^2)}\approx 30$~MeV.}
\begin{equation}
\chi^2(x)=\left(\frac{x-12.63}{0.45}\right)^2 + \left(\frac{x-83}{30}\right)^2, \label{eq:chi}
\end{equation}
which is minimized for $x=12.65$ with $\chi^2_0=5.50$, and the corresponding probability $p(\chi^2>\chi^2_0)=0.019<0.05$. Then we can conclude that the theoretical width $12.63\pm 0.45$~MeV of Case A is smaller than the LHCb result $83\pm21^{+21}_{-14}$~MeV at the 95\%CL. On the other hand, for the results of the Case B, we have,
\begin{equation}
\chi^2(x)=\left(\frac{x-31.3}{1.2}\right)^2 + \left(\frac{x-83}{30}\right)^2, \label{eq:chi}
\end{equation}
which is minimized for $x=31.4$ with $\chi^2_0=2.97$, and the corresponding probability $p(\chi^2>\chi^2_0)=0.085>0.05$. It implies that the value $31.3\pm1.2$~MeV of Case B is not significant smaller than the LHCb measurement from a statistical point of view.
 
 In addition, it is also easy to find that there are large discrepancies between  Case A and Case B, since the corresponding $\chi^2_0$ reads 212.2.
 Generally speaking, the different space wave functions would lead to different decay widths. Especially, if the overlap is near to the nodes of space wave functions, the decay width would strongly depend on the details of wave functions, and the small wave function difference could generate a large discrepancy of the decay width. The difference between the predictions in case A and case B provides a chance to distinguish two models.
Thus, if the small width of the $X(4140)$ is confirmed in future high-precision measurements, the $X(4140)$ could be explained as the charmonium state $\chi_{c1}(3P)$.
Indeed, the $B^+\to  J/\psi \phi K^+$ decay was investigated in Ref.~\cite{Wang:2017mrt}, where the $X(4140)$, with the small width $\Gamma=19$~MeV, and the molecular state $X(4160)$ were taken into account, and it was found that the low $J/\psi \phi$  invariant mass distributions were better described compared with the analysis in Refs.~~\cite{Aaij:2016iza,Aaij:2016nsc}， where only the $X(4140)$ resonance was considered.
Thus, the high-precision measurement about the $X(4140)$ width is necessary to shed light on its possible nature.

Studying the strong decay properties of the $\chi_{c0}(3P)$ and $\chi_{c2}(3P)$ states is also useful to search for those states, and understand the family of the charmonium states.
The decay widths of the $\chi_{c0}(3P)$ and $\chi_{c2}(3P)$ are tabulated in Table~\ref{tab:width4140}, and the initial mass dependences of the total widths are also shown in Fig.~\ref{fig:widthNRQM} and Fig.~\ref{fig:widthMGI}, respectively corresponding to the results of Case A and Case B. The total decay width of $\chi_{c0}(3P)$ is about $25\pm 3$~MeV for Case A with the predicted mass $4131\pm30$~MeV, and about $35\pm 5$~MeV for Case B with the predicted mass $4177\pm30$~MeV.
For the $\chi_{c2}(3P)$, the total decay width is predicted  to be about $35\pm 5$~MeV for Case A with the predicted mass $4208\pm30$~MeV, and about $43\pm 5$~MeV for Case B with the predicted mass $4213\pm30$~MeV. In the energies region of $4100 \sim 4250$~\cite{PDG2018}, there is one state $X(4160)$, with $M=4156^{+29}_{-25}$~MeV and $I^G(J^{PC}=?^?(?^{??})$, but with $\Gamma=139^{+110}_{-60}$~MeV, which is much larger than the predicted total widths of the $\chi_{cJ}(3P)$. Indeed, among the different interpretations of the $X(4160)$, the $D_s^*\bar{D}^*_s$ molecular nature has been widely studied in Refs.~\cite{Wang:2017mrt,Molina:2009ct,Wang:2018djr,Torres:2016oyz}.

\begin{table*}[htpb]
\begin{center}
\caption{\label{tab:width4140} Decay widths of the ${\chi}_{c0}(3P)$, ${\chi}_{c1}(3P)$ and ${\chi}_{c2}(3P)$ states (in MeV). The mass of the ${\chi}_{c1}(3P)$ is taken to be the one of the $X(4140)$, and the masses of the ${\chi}_{c0}(3P)$ and ${\chi}_{c2}(3P)$ are taken from the Table~\ref{tab:ccbar}, respectively for Case A and Case B.}
\begin{tabular}{c|c|c|c|c}
\hline\hline
 State                &Channel                        &Mode                 &$\Gamma$ (Case A)  &$\Gamma$ (Case B)\\
\hline
${\chi}_{c0}(3P)$     &$0^+\rightarrow 0^-0^-$           &$D\bar{D}$           &10.58          &0.22    \\
    $$                &$$                             &$D^+_sD^-_s$           &0.37            &1.87    \\
    $$                &$0^+\rightarrow 1^-1^-$            &$D^*\bar{D}^*$       &16.28           &35.95  \\
Total Width           &$$                             &$$                    &27.23          &38.03  \\
\hline
${\chi}_{c1}(3P)$     &$1^+\rightarrow 0^-1^-$          &$D\bar{D}^*$         &4.54           &14.48     \\
    $$                &$$                             &$D_s\bar{D}_s^*$     &1.23            &0.70    \\
    $$                &$1^+\rightarrow 1^-1^-$           &$D^*\bar{D}^*$       &6.86           &16.17   \\
Total Width           &$$                             &$$                    &12.63          &31.34    \\
\hline
${\chi}_{c2}(3P)$     &$2^+\rightarrow 0^-0^-$           &$D\bar{D}$           &7.71            &8.79      \\
    $$                &$$                             &$D^+_sD^-_s$           &0.63            &0.10 \\
    $$                &$2^+\rightarrow 0^-1^-$           &$D\bar{D}^*$         &20.04            &11.34      \\
    $$                &$$                             &$D_s\bar{D}_s^*$     &0.17            &0.13     \\
    $$                &$2^+\rightarrow 1^-1^-$           &$D^*\bar{D}^*$       &11.33            &26.87    \\
Total Width          &$$                             &$$                    &39.89         &47.23    \\
\hline\hline
\end{tabular}
\end{center}
\end{table*}

Finally, we would like to discuss about the uncertainties of the charmonium spectrum. We have extracted the wave functions from the MGI and MNRQM, and have not taken into account the error of the parameters. Since the errors of the established charmonium state are very small, we could expect that the errors of the predictions of these two models are also very small. The more important is that, the information of its quantum numbers $J^{PC}=1^{++}$ and the mass $4146.8\pm2.4$ are enough for one to obtain its possible assignment, and then we could calculate the decay width with this assignment.

\section{SUMMARY}
\label{sec:summary}
We have investigated the strong decay properties of the $X(4140)$ with the assignment of the $\chi_{c1}(3P)$ states in the $^3P_0$ model, where the modified non-relativistic quark model (Case A) and the modified Godfrey-Isgur relativistic quark model (Case B), both taking into account the screening effect, are used to extract the wave functions for the mesons. The only free parameter $\gamma$, the strength of the quark-antiquark pair created from the vacuum, is taken by fitting to the widths of the four well established charmonium states,   $\psi(3770)$  ($1^3D_1$), $\psi(4040)$ ($3^3S_1$), $\psi(4160)$ ($2^3D_1$), and $\chi_{c2}(2P)$.

The total decay width of the $\chi_{c1}(3P)$ is predicted to be $12.63\pm 0.45$~MeV for Case A, and $31.3\pm 1.2$~MeV for Case B, both of which support a narrow width for the $X(4140)$ resonance. Thus, we conclude that, the $X(4140)$, with a small width, could be explained as the charmonium state $\chi_{c1}(3P)$, and the high-precision measurement about the $X(4140)$ could shed light on its nature.

We have also performed a simple $\chi^2$ study, which shows that the value $12.63\pm 0.45$~MeV of Case A is smaller than the LHCb measurement at the 95\%CL, and the one $31.3\pm 1.2$~MeV of Case B
is not significant smaller than the LHCb measurement from a statistical point of view.

We also show the strong decay properties of $\chi_{c0}(3P)$ and $\chi_{c2}(3P)$, and the total widths of the $\chi_{c0}(3P)$ and $\chi_{c2}(3P)$ are predicted be about $20\sim 40$~MeV  and $30\sim 50$~MeV, respectively. By comparing with the width of the $X(4160)$,  we find it is difficult to interpretation the $X(4160)$ as the charmonium states $\chi_{cJ}(3P)$.

\section*{Acknowledgements}
This work is partly supported by the National Natural Science Foundation of China under Grant No. 11505158, the Key Research Projects of Henan Higher Education Institutions (No. 20A140027), and the Academic Improvement Project of Zhengzhou University.

\end{document}